\documentclass[superscriptaddress,prd,aps,showpacs,nofootinbib,showkeywords,eqsecnum,preprint]{revtex4-1}

\usepackage{graphicx,color,amsmath,amsxtra}
\usepackage{epsf}
\usepackage{amssymb}
\usepackage{enumerate}
\usepackage{hhline}
\usepackage{array}
\usepackage{tabularx}
\usepackage[unicode]{hyperref}
\usepackage{graphicx}                
\usepackage{epstopdf}

\begin{document}
\pagestyle{myheadings}

\title{Anisotropic power-law inflation for a generalized model of two scalar and two vector fields}
\author{Tuan Q. Do }
\email{tuan.doquoc@phenikaa-uni.edu.vn}
\affiliation{Phenikaa Institute for Advanced Study, Phenikaa University, Hanoi 12116, Vietnam}
\affiliation{Faculty of Basic Sciences, Phenikaa University, Hanoi 12116, Vietnam}
\author{W. F. Kao}
\email{homegore09@nycu.edu.tw}
\affiliation{
Institute of Physics, National Yang Ming Chiao Tung University, Hsin Chu 30010, Taiwan
}
\date{\today} 
\begin{abstract}
Cosmological implication of a generalized model of two scalar and two vector fields, in which both scalar fields are non-minimally coupled to each vector field, is studied in this paper. In particular, we will seek a set of new anisotropic power-law inflationary solutions to this model. Additionally, the stability of the obtained solutions will be examined by using the dynamical system approach. As a result, we will show that this set of solutions turns out to be stable and attractive during the inflationary phase as expected due to the existence of the unusual couplings between two scalar and two vector fields. Notably, we will point out that the existence of phantom field will lead to an instability of the corresponding anisotropic power-law inflation.
\end{abstract}

\maketitle
\section{Introduction} \label{intro}
In modern cosmology, the cosmological principle stating that our universe is simply homogeneous and isotropic on large scales has been an underlying assumption of not only cosmic inflation models associated with the early universe \cite{Starobinsky:1980te,Guth:1980zm,Linde:1981mu,Linde:1983gd} but also of the concordance  $\Lambda$CDM model associated with the cosmic acceleration of the late time universe \cite{Frieman:2008sn}. Remarkably, most of observational data have turned out to be consistent with this principle for several decades. 

However, recent high resolution observations and deep analysis have provided important hints for a refinement of the cosmological principle \cite{Komatsu:2010fb,WMAP,Ade:2015hxq,Akrami:2019bkn,Schwarz:2015cma,FLRW,Krishnan:2021dyb}. This is due to the fact that a number of observational anomalies and tensions have increased over time. For example, some unavoidable anomalies of the cosmic microwave background radiations (CMB), e.g., the hemispherical asymmetry and the cold spot, which have been revealed by the Wilkinson Microwave Anisotropy Probe satellite (WMAP) \cite{Komatsu:2010fb,WMAP} and then verified by the Planck satellite \cite{Ade:2015hxq,Akrami:2019bkn}, turn out to be inconsistent with theoretical predictions of the standard inflationary models \cite{Schwarz:2015cma}, whose backbone is the Friedmann-Lemaitre-Robertson-Walker (FLRW) spacetime \cite{FLRW} obeying the cosmological principle. In addition to these anomalies, a recent study has suggested that the so-called Hubble tension might be signal for a breakdown in FLRW cosmology \cite{Krishnan:2021dyb}. 

It is worth noting that the Hubble tension has emerged as one of the hottest subjects in cosmology recently due to an unavoidable gap between the values of Hubble constant determined from different observations of two phases of the universe, one is the early time CMB observations and the other is the late time supernova observations \cite{Riess:2016jrr,Riess:2019cxk}. For interesting reviews on this tension, see Refs.  \cite{DiValentino:2021izs,Perivolaropoulos:2021jda,Abdalla:2022yfr}. Remarkably, a number of possible resolutions of the Hubble tension have been proposed in literature, e.g., see Refs. \cite{Poulin:2018cxd,Marra:2021fvf,Alestas:2021nmi,Alestas:2020zol}. In particular, some people have imagined that the existence of early dark energy, which acts as a cosmological constant at early times and then dilutes away like radiation or faster at later times, can help us to solve the Hubble tension  \cite{Poulin:2018cxd}. On the other hand, some other people have  thought that the Hubble tension might be more of an astrophysical problem than a cosmological one \cite{Marra:2021fvf}. In particular, they have assumed that the Hubble tension is due to  a gap between two values of the absolute magnitude $M$ of type Ia supernovae, one is calibrated using Cepheid stars in the redshift range $0<z<0.01$ (distance ladder calibration) and the other is  calibrated using
the recombination sound horizon (inverse distance ladder calibration) for $0.01 < z<z_{rec}$ \cite{Alestas:2021nmi}. Therefore, a transition of $M$ at  a redshift $z_t \simeq 0.01$ might be a reasonable resolution to the Hubble tension \cite{Marra:2021fvf,Alestas:2021nmi,Alestas:2020zol}. 

These remarkable points would lead us to a possible scenario that the early universe may not be compatible with the cosmological principle. In other words, the early universe may be slightly inhomogeneous and/or anisotropic on large scales. One may therefore address a related question that what if the cosmological principle does not hold in the late time universe. A recent analysis based on supernova observations has called the cosmological principle into question when it has arrived at an important result that the late time universe may be anisotropic \cite{Colin:2018ghy}. The breakdown of the cosmological principle in the late time universe has also been pointed out in the follow-up studies based on quasar observations \cite{Secrest:2020has} or supernova observations \cite{Krishnan:2021jmh}. Other important works indicating the violation of the cosmological principle can be found in an interesting review \cite{Aluri:2022hzs}. 

Along with the cosmological principle, there has existed the well-known cosmic no-hair conjecture proposed by Hawking and his colleagues long time ago, which is also concerned with the homogeneity and isotropy of the universe \cite{Gibbons:1977mu,Hawking:1981fz}. In particular, the late time universe should, according to this conjecture, obey the cosmological principle, no matter how its initial states were, which could even be inhomogeneous and/or anisotropic. This conjecture has received a lot of attentions since the Wald's seminal proof for the Bianchi spacetimes with the presence of a cosmological constant $\Lambda$ \cite{Wald:1983ky,Barrow:1987ia,Mijic:1987bq,Kitada:1991ih,barrow05,Maleknejad:2012as,Kleban:2016sqm,East:2015ggf,Carroll:2017kjo,MW0,MW}. Remarkably, Starobinsky and other people have pointed out  that the cosmic no-hair conjecture would only be valid inside of the future event horizon \cite{Starobinsky:1982mr,Muller:1989rp,Barrow:1984zz,Jensen:1986nf,SteinSchabes:1986sy}.

Recently, this conjecture has been called in question by many studies. On the observational side, the prediction of this conjecture seems to be inconsistent with the investigations based on late time observational data in Refs. \cite{Colin:2018ghy,Secrest:2020has,Krishnan:2021jmh}. On the theoretical side, an interesting counterexample to this conjecture has been found by Soda and his colleagues  \cite{MW0,MW}. In particular, they have successfully derived a set of stable and attractive Bianchi type I inflationary solutions to a supergravity-motivated model involving an unusual non-minimal coupling between scalar and vector fields as $f^2(\phi)F_{\mu\nu}F^{\mu\nu}$ \cite{MW}. It turns out that the Bianchi type I metric is the simplest type among the Bianchi metrics, which are homogeneous but anisotropic spacetimes \cite{Ellis:1968vb}. The stability of Bianchi type I inflationary solution would therefore lead to the abundance of spatial (small) anisotropies after the inflationary phase, which might still exist in the late time universe, in accord with recent observational investigations  \cite{Colin:2018ghy,Secrest:2020has,Krishnan:2021jmh}. It turns out that the Kanno-Soda-Watanabe (KSW) model \cite{MW0} has received a lot of attentions \cite{Soda:2012zm,Maleknejad:2012fw}. A number of its extensions have been constructed to see if  the cosmic no-hair conjecture is still disfavored \cite{Emami:2010rm,Murata:2011wv,Hervik:2011xm,Do:2011zza,Do:2011zz,Yamamoto:2012tq,Thorsrud:2012mu,Maeda:2012eg,Ohashi:2013mka,Do:2016ofi,Tirandari:2017nzy,Do:2017rva,Ito:2017bnn,Do:2020hjf,Do:2021lyf,Chen:2021nkf,Nguyen:2021emx,Gao:2021qwl,Do:2021pqk,Chen:2022ccf,Goodarzi:2022wli,Kanno:2022flo}. It turns out that the conjecture has been shown to be extensively violated in many scenarios of these extensions. The main reason for this violation is due to the appearance of the non-constant gauge kinetic function $f(\phi)$ in the coupling, $f^2(\phi)F_{\mu\nu}F^{\mu\nu}$, which eliminates the conformal invariance of vector field and therefore keeps the vector field away from being rapidly diluted during the inflationary phase, leading to the appearance of stable small spatial anisotropies. 

In this paper, we would like to generalize our recent work \cite{Do:2021lyf}, which is a multi-field extension of the KSW model, by proposing a more general scenario that two scalar fields can be simultaneously non-minimally coupled to each vector field. As a result, a set of anisotropic solutions of this model can be easily reduced to that found in Ref. \cite{Do:2021lyf} by taking suitable limits. Furthermore, this set of solutions will be shown to be attractive and stable, similar to many counterexamples found in the previous papers. This result emphasizes that the cosmic no-hair conjecture is still violated in the present model of multiple scalar and vector fields. However, we will point out that if one of two scalar fields is flipped from canonical to phantom then at least one unstable mode will emerge, in accord with the prediction of the cosmic no-hair conjecture.

In summary, this paper will be organized as follows: (i) Its brief introduction has been written in Sec. \ref{intro}. (ii) Basic setup of the proposed model along with its set of new anisotropic power-law inflationary solutions will be presented in Sec. \ref{sec2}. (iii) Stability analysis of the obtained solutions will be investigated in Sec. \ref{sec3}. (iv) Finally, concluding remarks will be given in Sec. \ref{final}.
\section{The model and its anisotropic power-law solution} \label{sec2}
\subsection{Basic setup}
Motivated by the result obtained in our previous studies \cite{Do:2011zza,Do:2021lyf}, we would like to consider a more general action of two scalar and two vector fields in four dimensions such as\\
\begin{align} \label{Einstein-general}
S=  \int d^4 x \sqrt{- g}& \left[ \frac{ R}{2} - \frac{1}{2} \left( \partial_\mu \phi \right) \left(\partial^\mu \phi \right)  - \frac{1}{2}  \left( \partial_\mu \psi \right) \left( \partial^\mu \psi\right) -V(\phi,\psi) \right. \nonumber\\
&\left.- \frac{1}{4} {f_1^2} e^{a \phi + b\psi} { F}^{1\mu\nu} { F} ^1_{\mu\nu}  - \frac{1}{4} {f_2^2} e^{c\phi +d \psi } {F}^{2\mu\nu} {F}^2_{\mu\nu}    \right],
\end{align}
where $M_p=1$, while $f_1$, $f_2$, $a$, $b$, $c$, and $d$ are all arbitrary constants. Furthermore, $\phi$ and $\psi$ both are canonical scalar fields, while $F^1_{\mu\nu} \equiv \partial_\mu A^1_\nu -\partial_\nu A^1_\mu$ and $F^2_{\mu\nu}\equiv \partial_\mu A^2_\nu -\partial_\nu A^2_\mu$ are the field strength of two vector fields, $A^1_\mu$ and $A^2_\mu$, respectively. In this model, both scalar fields are assumed to simultaneously non-minimally couple to each vector field. Moreover, they are also assumed to be homogeneous and therefore depend only on the cosmic time $t$. It appears that if one of two vector fields vanishes along with an assumption that $\psi$ is nothing but a phantom field then the above action will reduce to that proposed in Ref. \cite{Do:2011zza}. On the other hand, if we set $b=c=0$ then the action proposed in Ref. \cite{Do:2021lyf} will be easily recovered from the above action. It should be noted that multi-vector-field extensions of the KSW model can be found in Ref. \cite{Yamamoto:2012tq}.

 Our next goal is to investigate whether the generalized action \eqref{Einstein-general} admits a set of new stable and attractive anisotropic inflationary solutions. We must, therefore, define the corresponding field equations from this action using the principle of least action.  As a result, the generalized Einstein field equation of this model is defined to be
\begin{align} \label{Einstein-field}
& R_{\mu\nu} -\frac{1}{2}Rg_{\mu\nu } -\partial_\mu \phi \partial_\nu \phi  - \partial_\mu \psi \partial_\nu \psi \nonumber\\
&+\frac{1}{2} g_{\mu\nu} \left[\partial_\sigma \phi \partial^\sigma \phi +\partial_\sigma \psi \partial^\sigma \psi +2V+\frac{1}{2}f_1^2 e^{a \phi + b\psi} (F^1)^2+ \frac{1}{2} f_2^2 e^{c\phi +d \psi } ({ F^2})^2  \right] \nonumber\\
&-f_1^2 e^{a \phi + b\psi} F^1_{\mu\gamma}F^1_\nu{}^{\gamma} -f_2^2 e^{c\phi +d \psi }  {F}^2_{\mu\gamma} { F}^2_\nu{}^{\gamma}=0.
\end{align}
In addition, the field equations of two vector fields, $A^1_\mu$ and ${A}^2_\mu$, are given by 
\begin{align} \label{vector-field-1}
\partial_\mu \left[\sqrt{-g}   e^{a \phi + b\psi} F^{1\mu\nu}  \right] &=0,\\
\label{vector-field-2}
\partial_\mu \left[\sqrt{-g}  e^{c\phi +d \psi } { F}^{2\mu\nu} \right] &=0,
\end{align}
respectively. Finally, the field equations of two scalar fields, $\phi$ and $\psi$, read \begin{align} \label{phi-1-scalar-field}
\square \phi -\partial_{\phi} V-\frac{a}{4}  f_1^2  e^{a \phi + b\psi}  (F^1)^2 -\frac{c}{4}  f_2^2  e^{c \phi + d\psi}  (F^2)^2 &=0,\\
\label{phi-2-scalar-field}
 \square \psi -\partial_{\psi} V-\frac{b}{4} f_1^2 e^{a\phi +b \psi } (F^1)^2-\frac{d}{4} f_2^2 e^{c\phi +d \psi } (F^2)^2 &=0,
\end{align}
respectively. Here, it is understood that $\partial_{\phi}V  \equiv \partial V/\partial \phi $, $\partial_{\psi} V \equiv \partial V/\partial \psi $, and $\square \equiv \frac{1}{\sqrt{-g}} \partial_\mu \left(\sqrt{-g} \partial^\mu \right)$ is the d’Alembert operator. 
In order to investigate a set of new anisotropic inflationary solutions, we will consider the simplest Bianchi spacetime, i.e., the Bianchi type I metric  \cite{MW0,MW,Ellis:1968vb}
\begin{equation} \label{metric}
ds^2 =-dt^2 +e^{ 2\alpha(t) -4\sigma(t)} dx^2 +e^{ 2\alpha(t) +2\sigma(t)}\left(dy^2+dz^2 \right).
\end{equation}
It should be noted that the following constraint, $|\alpha(t)/\sigma(t)| \gg 1 $, should be satisfied during an inflationary phase since $\alpha(t)$ and $\sigma(t)$ characterise the spatial isotropy and anisotropy of spacetime, respectively. Of course, one can ask if inhomogeneous and anisotropic spacetimes could be relevant to the KSW model as well as its multi-field extensions like this model. However, it would become very complicated if we consider such spacetimes. For now, we just prefer the Bianchi type I metric, which is the straightforward generalization of the FLRW one,  for simplicity, following many previous papers \cite{MW0,MW,Do:2011zza,Do:2021lyf}. Additionally, by regarding the same Bianchi type I metric, we could observe the effect of extra fields on the scale factors when comparing the obtained solutions with those derived in the previous papers \cite{MW,Do:2011zza,Do:2021lyf}.

 Similar to Refs. \cite{MW0,MW,Do:2021lyf}, the setup of the vector fields, $A^1_\mu$  and $A^2_\mu$, will be chosen as  $A^1_\mu   = \left( {0,A^1_x \left( t \right),0,0} \right)$ and  $A^2_\mu   = \left( {0,A^2_x \left( t \right),0,0} \right)$  due to the $y-z$ rotational symmetry of the Bianchi metric. As a result, Eqs. \eqref{vector-field-1} and \eqref{vector-field-2}, admit their non-trivial solutions,
\begin{align}
{\dot A}^1_x & =p_A   e^{-a\phi-b\psi-\alpha-4\sigma},\\
{\dot  A}^2_x &=q_A e^{-c\phi-d\psi-\alpha-4\sigma},
\end{align}
respectively, where $p_A$ and $q_A$ are constants of integration. Here $\dot A^1_x \equiv dA^1_x/dt$ and $\dot A^2_x \equiv dA^2_x/dt$. It is clear that if the gauge kinetic functions, $e^{a\phi+b\psi}$ and $e^{c\phi+d\psi}$, are absent then ${{\dot A}^{1}_x ~({\dot A}^{2}_x) \sim e^{-\alpha}}$ and therefore ${F^{1}_{\mu\nu}F^{1\mu\nu} ~(F^{2}_{\mu\nu}F^{2\mu\nu})  \sim e^{-4\alpha}}$. As $\alpha$ increases quickly during the inflationary phase then the vector fields will decay exponentially to zero. The existence of the gauge kinetic functions will help us to balance $e^{-4\alpha}$ such that the vector fields will not be rapidly diluted during the inflationary phase. Of course, the contribution of the vector fields in the field equations should be in harmony with the existence of small spatial anisotropies of spacetime during the inflationary phase. The breaking of rotational invariance of the vector fields has been made in accord with the existence of spatial anisotropies of the Bianchi type I metric shown above. Consequently, the corresponding small spatial hairs (a.k.a. spatial anisotropies) of the present model will be derived in the next subsection. More interestingly, non-trivial imprints of non-vanishing component of vector field(s) on the CMB may be relevant to future observations as pointed out in the previous works for one scalar field non-minimally coupled to one vector field  \cite{Watanabe:2010bu,Do:2020ler}. In this paper, however, we focus only on the stability of anisotropic inflationary models. Due to complicated and lengthy calculations, CMB imprints of multi-field extensions of the KSW model will be investigated systematically in the future and will be presented elsewhere. 

Thanks to the above solutions and setup,  the field equations \eqref{Einstein-field}, \eqref{phi-1-scalar-field}, and \eqref{phi-2-scalar-field} now take their explicit expressions,
\begin{align} \label{field-equation-1}
\dot\alpha^2 &= \dot\sigma^2 +\frac{1}{3} \left[ \frac{\dot\phi^2}{2}+ \frac{\dot\psi^2}{2}+V +\frac{1}{2} \left(p_A^2 f_1^{2} e^{-a\phi-b\psi}+q_A^2 f_2^{2} e^{-c\phi-d\psi}  \right)e^{-4\alpha-4\sigma} \right],\\
\label{field-equation-2}
\ddot\alpha&=-3\dot\alpha^2 +V +\frac{1}{6} \left(p_A^2 f_1^{2} e^{-a\phi-b\psi}+q_A^2 f_2^{2} e^{-c\phi-d\psi}  \right)e^{-4\alpha-4\sigma},\\
\label{field-equation-3}
\ddot\sigma&=-3\dot\alpha \dot\sigma +\frac{1}{3} \left(p_A^2 f_1^{2} e^{-a\phi-b\psi}+q_A^2 f_2^{2} e^{-c\phi-d\psi}  \right)e^{-4\alpha-4\sigma} ,\\
\label{field-equation-4}
\ddot\phi&=-3\dot\alpha \dot\phi -\partial_{\phi} V +\frac{a}{2}  p_A^2 f_1^{2}e^{-a\phi-b\psi-4\alpha-4\sigma} + \frac{c}{2}  q_A^2 f_2^{2}e^{-c\phi-d\psi-4\alpha-4\sigma},\\
\label{field-equation-5}
\ddot\psi&=-3\dot\alpha \dot\psi -  \partial_{\psi} V+\frac{b}{2}  p_A^2 f_1^{2}e^{-a\phi-b\psi-4\alpha-4\sigma}  + \frac{d}{2}q_A^2 f_2^{2} e^{-c\phi-d\psi-4\alpha-4\sigma},
\end{align} 
respectively. Here, $\ddot\alpha \equiv d^2\alpha/ dt^2$ and so on.
Among three Einstein field equations shown above, the first equation, Eq. \eqref{field-equation-1}, is the Friedmann equation acting as a constraint field equation, which any obtained solutions must satisfy. The second one, Eq. \eqref{field-equation-2}, will encode the evolution of the spatial isotropy $\alpha$. The last one,  Eq. \eqref{field-equation-3}, which is restricted to the Bianchi type I spacetime, will control the evolution of the spatial anisotropy. These equations are the key equations describing the expansion of the early universe. It is apparent that the last term in the right hand side of Eq. \eqref{field-equation-3} plays an important role in maintaining the appearance of small spatial anisotropy during the inflationary phase. Without it, we can only have isotropic inflation with $\dot\sigma\sim e^{-3\alpha} \sim 0$ as $\alpha \gg 1$.
\subsection{Exact anisotropic power-law solution}
Given the derived field equations, we now would like to seek their exact anisotropic power-law solutions, following the previous works \cite{MW,Do:2011zza,Do:2021lyf,Do:2021pqk}. To complete this task, we will consider an exponential potential $V$ such as \cite{Do:2011zza,Do:2021lyf}
\begin{equation} \label{exponential-potential}
V(\phi,\psi)=V_{01}e^{\lambda_1 \phi}+V_{02}e^{\lambda_2 \psi},
\end{equation}
along with the following compatible ansatz given by
\begin{align} \label{ansatz}
\alpha (t)= \zeta \log t,~\sigma(t) = \eta \log t, \nonumber\\
\phi(t) = \xi_1 \log t +\phi_{0}, ~\psi(t) = \xi_2 \log t +\psi_{0}.
\end{align}
Here $\phi_{0}$, $\psi_{0}$, $\xi_i$, $V_{0i}$, and $\lambda_i$ are non-zero parameters. It turns out that the scale factors of the Bianchi type I metric shown in Eq. \eqref{metric} now become power-law functions of cosmic time $t$,
\begin{equation}
e^{ 2\alpha(t) -4\sigma(t)}  = t^{2\zeta-4\eta}, \quad e^{ 2\alpha(t) +2\sigma(t)} = t^{2\zeta+2\eta}.
\end{equation}
For an inflationary universe, it requires that $\zeta-2\eta \gg 1$ as well as $\zeta+\eta\gg 1$ with $\zeta \gg \eta$ \cite{MW}.
Additionally, the choice of exponential potentials shown in Eq. \eqref{exponential-potential} together with the above ansatz \eqref{ansatz} will also lead to power-law functions of cosmic time $t$,
\begin{equation}
V(\phi,\psi) = V_{01} e^{\lambda_1 \phi_0}t^{\lambda_1 \xi_1}+V_{02} e^{\lambda_2 \psi_0}t^{\lambda_2 \xi_2}.
\end{equation}
Similar thing also happens to the gauge kinetic functions, i.e., 
\begin{equation}
e^{a\phi +b\psi} = e^{a\phi_0+b\psi_0} t ^{a\xi_1 +b \xi_2}, ~e^{c\phi +d\psi} = e^{c\phi_0+d\psi_0} t ^{c\xi_1 +d \xi_2}.
\end{equation}

It is noted that the choice of exponential potentials and gauge kinetic functions has been made with the expectation of the existence of power-law inflationary solutions. Furthermore, it helps us to easily recover the previous power-law solutions published in Refs. \cite{MW,Do:2011zza,Do:2021lyf}. It also helps us to see the contribution of additional vector field to the value of scale factors of spacetime when comparing the generalized solutions derived in this paper with the previous ones. It should be noted that the type of power-law inflation considered in this paper is different from that investigated in cosmological models including power-law potentials, e.g., see Refs. \cite{Keskin:2021zct,Luciano:2023roh}, in which power-law potentials show up in the context of extended cosmology based on deformed entropy-area laws. However, power-law potential(s) would be relevant to an anisotropic hyperbolic inflation model proposed in Ref. \cite{Do:2021pqk}.

As a result, the differential field equations \eqref{field-equation-1}, \eqref{field-equation-2}, \eqref{field-equation-3}, \eqref{field-equation-4}, and \eqref{field-equation-5} give us the following set of algebraic equations,
\begin{align}
\label{algebraic-1}
\zeta^2 &= \eta^2 +\frac{1}{3} \left[\frac{\xi_1^2}{2}+ \frac{\xi_2^2}{2}+ u_1+u_2 +\frac{1}{2} \left(v_1 +v_2 \right) \right],\\
\label{algebraic-2}
-\zeta&= -3\zeta^2 +u_1+u_2 +\frac{1}{6} \left(v_1+v_2 \right),\\
\label{algebraic-3}
-\eta&= -3\zeta \eta +\frac{1}{3} \left(v_1+v_2  \right),\\
\label{algebraic-4}
-\xi_1 &= -3\zeta \xi_1 -\lambda_1 u_1 + \frac{a}{2}  v_1+\frac{c}{2}v_2,\\
\label{algebraic-5}
-\xi_2 &= -3\zeta \xi_2 - \lambda_2 u_2 + \frac{b}{2} v_1+ \frac{d}{2}  v_2,
\end{align}
thanks to the corresponding constraint equations,
\begin{align}
\label{constraint-1}
\lambda_1 \xi_1 &= -2,\\
\label{constraint-2}
\lambda_2 \xi_2 &= -2,\\
\label{constraint-3}
2\zeta+2\eta+\frac{a}{2} \xi_1 +\frac{b}{2}\xi_2 & =1,\\
\label{constraint-4}
2\zeta+2\eta+\frac{c}{2}\xi_1+\frac{d}{2} \xi_2 & =1,
\end{align}
by which the existence of the cosmic time $t$ in the field equations can be ignored since all terms in the mentioned field equations turn out to be directly proportional to $t^{-2}$. For convenience, we have introduced four additional variables $u_i$ and $v_i$ ($i=1-2$), whose definitions are given by
\begin{align}
u_1&=V_{01} e^{\lambda_1 \phi_{0}},\\
u_2&=V_{02} e^{\lambda_2 \psi_{0}},\\
v_1&=  p_A^2 f_{1}^{2} e^{-a \phi_{0} -b\psi_{0}},\\
v_2&=  q_A^2 f_{2}^{2} e^{-c\phi_{0} -d \psi_{0}}.
\end{align}
It turns out from two  Eqs. \eqref{constraint-1} and \eqref{constraint-2} that
\begin{align} 
\lambda_1 \xi_1= \lambda_2 \xi_2 ,
\end{align}
while the  other two, Eqs. \eqref{constraint-3} and \eqref{constraint-4}, imply
\begin{align}
\frac{a}{\lambda_1} +\frac{b}{\lambda_2} = \frac{c}{\lambda_1}+\frac{d}{\lambda_2} =\kappa_1,
\end{align}
with $\kappa_1$ is an additional constant. Furthermore, the last relation  yields
\begin{equation}
a\lambda_2 +b\lambda_1 = c\lambda_2 +d \lambda_1 =\kappa_2,
\end{equation}
with $\kappa_2$ is an another additional constant. This indicates that if all $a$, $b$, $c$, and $d$ are fixed then $\lambda_1$ and $\lambda_2$ are related to each other such as
\begin{equation}
\lambda_2 = \frac{d-b}{a-c}\lambda_1.
\end{equation}
If we assume $\lambda_1$ and $\lambda_2$ are both positive definite then the inequality $d>b$ will require another inequality $a>c$, or vice versa. 
Interestingly, it appears that $\lambda_1$ and $\lambda_2$ will be independent of each other if  $a=c$ along with $b=d$. 

Multiplying both sides of Eq. \eqref{algebraic-4} with $\xi_1$, both sides of Eq. \eqref{algebraic-5} with $\xi_2$, then combining them will lead to
\begin{equation} \label{equation-of-u-v}
-\left(3\zeta-1\right) \left(\frac{4}{\lambda_1^2} +\frac{4}{\lambda_2^2} \right) +2u-\kappa_1 v =0,
\end{equation}
where we have used the constraints shown in Eqs. \eqref{constraint-1} and \eqref{constraint-2} and  defined new variables \cite{Do:2021lyf}
\begin{align}
u&= u_1+u_2,\\
v&=v_1+v_2.
\end{align}
Furthermore, these two variables can be determined in terms of $\zeta$ and $\eta$, according to two Eqs. \eqref{algebraic-2} and \eqref{algebraic-3}, as
\begin{align}
u&= \zeta \left(3\zeta-1 \right) -\frac{v}{6},\\
v&=3\eta \left(3\zeta-1 \right).
\end{align}
Additionally, we have from Eq. \eqref{constraint-3} or Eq. \eqref{constraint-4} that
\begin{equation}
\eta =-\zeta+\frac{\kappa_1}{2} +\frac{1}{2}.
\end{equation}
Finally, inserting these results into either Eq. \eqref{algebraic-1} or Eq. \eqref{equation-of-u-v} leads to an equation of $\zeta$ such as
\begin{equation}
6 \lambda_1^2 \lambda_2^2 \left(\kappa_1+1 \right) \zeta  -\lambda_1^2\lambda_2^2 \left(3\kappa_1^2 +4\kappa_1 +1 \right)-8\lambda_1^2-8\lambda_2^2=0.
\end{equation}
Consequently, a non-trivial solution of $\zeta$ turns out to be
\begin{equation} \label{solution-zeta}
\zeta =\frac{\lambda_1^2\lambda_2^2 \left(3\kappa_1^2 +4\kappa_1 +1 \right)+8\lambda_1^2+8\lambda_2^2  }{6\lambda_1^2 \lambda_2^2 \left(\kappa_1+1 \right) }.
\end{equation}
Thanks to this solution, we are able to define the corresponding value of $\eta$, $u$, and $v$ as
\begin{align}\label{solution-eta}
\eta & = \frac{ \lambda_1^2\lambda_2^2 \left(\kappa_1 +1 \right)-4\lambda_1^2-4 \lambda_2^2}{3\lambda_1^2 \lambda_2^2 \left(\kappa_1+1 \right) },\\
u&= \frac{\left[ \lambda_1^2 \lambda_2^2 \left( 3\kappa_1^2+2\kappa_1 -1 \right)+8\lambda_1^2 +8\lambda_2^2 \right] \left[\lambda_1^2 \lambda_2^2 \kappa_1 \left(\kappa_1+1 \right) +4\lambda_1^2+4\lambda_2^2 \right] }{4\lambda_1^4\lambda_2^4 \left(\kappa_1+1 \right)^2},\\
v&= \frac{\left[ \lambda_1^2 \lambda_2^2 \left( 3\kappa_1^2+2\kappa_1 -1 \right)+8\lambda_1^2 +8\lambda_2^2 \right] \left[\lambda_1^2 \lambda_2^2  \left(\kappa_1+1 \right) -4\lambda_1^2-4\lambda_2^2 \right] }{2\lambda_1^4\lambda_2^4 \left(\kappa_1+1 \right)^2}.
\end{align}
As a result, the corresponding anisotropy parameter is given by
\begin{equation}
{ \frac{\Sigma}{H} \equiv \frac{\dot\sigma}{\dot\alpha}}= \frac{\eta}{\zeta} =\frac{2\left[\lambda_1^2\lambda_2^2 \left(\kappa_1 +1 \right)-4\lambda_1^2-4 \lambda_2^2\right]}{\lambda_1^2\lambda_2^2 \left(3\kappa_1^2 +4\kappa_1 +1 \right)+8\lambda_1^2+8\lambda_2^2 },
\end{equation}
which can be related to the slow-roll parameter, which is defined as $\epsilon \equiv -\dot H/H^2 $,  as follows
\begin{equation}
\frac{\Sigma}{H}  =\frac{1}{3}I\epsilon,
\end{equation}
where $I = 3\eta$ and $H \equiv \dot\alpha$ being the Hubble parameter. It turns out that if $\kappa_1 \gg 1$ then
\begin{align}
\zeta & \simeq \frac{\kappa_1}{2} +\frac{4}{3\kappa_2} \left(\frac{\lambda_1}{\lambda_2}+\frac{\lambda_2}{\lambda_1}\right) \gg 1,\\
\eta & \simeq \frac{1}{3} -\frac{4}{3\kappa_2} \left(\frac{\lambda_1}{\lambda_2}+\frac{\lambda_2}{\lambda_1}\right).
\end{align}
Furthermore, the constraint $|\eta| <1$ implies that the following one,
\begin{equation} \label{constraint-kappa-2}
\frac{4}{3\kappa_2} \left(\frac{\lambda_1}{\lambda_2}+\frac{\lambda_2}{\lambda_1}\right) \sim {\cal O}(0.1).
\end{equation}
If we assume $\lambda_1$ and $\lambda_2$ as positive parameters then the constraint shown in Eq. \eqref{constraint-kappa-2} will require that
\begin{equation}
\kappa_2 >1,
\end{equation}
because of the fact,
\begin{equation}
 \frac{\lambda_1}{\lambda_2}+\frac{\lambda_2}{\lambda_1} >1.
 \end{equation}
Consequently, it appears that
\begin{equation}
{\frac{\Sigma}{H}} \sim \kappa_1^{-1} \ll 1.
\end{equation}
 It is noted that if $a$, $b$, $c$, and $d \sim {\cal O}(10)$ then $| \lambda_1 | \ll 1$ and $|\lambda_2| \ll 1$ are required  to ensure that $\kappa_1 \gg 1$.  However, $\lambda_1$ and $\lambda_2$ cannot arbitrarily be much smaller than one due to the requirement $|\eta| <1$. Indeed, if $\kappa_2 \ll 1$ and $\lambda_1 \sim \lambda_2$ then $\eta$ will be large unexpectedly. For a case $\lambda_1 \sim \lambda_2$, we should make sure that $\kappa_2 $ is much larger than one. It turns out that a simple choice, in which $\lambda_1=0.1$, $\lambda_2 =0.2$, $a=80$, $b=c=40$, and $d=120$, will easily satisfy the inflationary requirement, $\zeta \gg 1$ and $|\eta|<1$. Another  simple example can be taken is that $\lambda_1=0.1$, $\lambda_2 =0.2$, $a=c=80$, and $b=d=40$. We will consider these two examples for further numerical calculations. 
 
One might wonder how we can recover solutions found in the previous models \cite{MW,Do:2021lyf}. It can be quite straightforward to do this recovering by taking suitable  limits.  First,  if we set $a=b=c=d=\rho$ as well as $\lambda_1 =\lambda_2 =\lambda$ then the above solution of $\zeta$ will be reduced to
\begin{equation}
\zeta = \frac{\lambda^2+8\lambda \rho +12\rho^2 +16}{6\lambda \left(\lambda+2\rho \right)},
\end{equation}
which is a non-trivial solution of an extended KSW model having two identical scalar and two identical vector fields  \cite{MW}. Second, if we turn off one scalar field as well as one vector field, e.g., $a=2\rho$, $\lambda_1= \lambda$, and $b=c=d=0$, $\lambda_2 \to \infty$, then we will obtain the following solution,
\begin{equation}
\zeta = \frac{\lambda^2+8\lambda \rho +12\rho^2 +8}{6\lambda \left(\lambda+2\rho \right)},
\end{equation}
which is nothing but that derived in the original KSW model \cite{MW}.
Third, if we take limits $b\to 0$ and $c\to 0$ altogether we will obtain the solution found in Ref. \cite{Do:2021lyf}, provided that $a=2\rho_1$ as well as $d=2\rho_2$. 

To end this section, we would like to note that our technique of solving field equations can be extended to a more general case of multi scalar and vector fields, in which the total number of fields is larger than four, even when all scalar fields are simultaneously non-minimally coupled to each vector field since we can always define $u=\sum_i u_i $ and $v=\sum_i v_i$. Very interestingly, we can show that the value of $\zeta$ will always remain as $\zeta \simeq \kappa_1 /2 \gg 1$. This result is similar to that of a less general scenario of multi scalar and vector fields, in which one scalar field is just non-minimally coupled to one vector \cite{Do:2021lyf}.
\section{Stability analysis} \label{sec3}
\subsection{Dynamical system and its anisotropic fixed point}
We have derived the set of new anisotropic power-law inflationary solutions to the proposed model. An important issue we should address for now is about its stability during the inflationary phase. It is noted that if the obtained anisotropic solution is confirmed to be stable against field perturbations, it will be an extra counterexample to the cosmic no-hair conjecture. Similar to the previous works \cite{MW,Do:2021lyf,Do:2021pqk}, we prefer using the dynamical system, which is the standard method to study the stability of cosmological systems in both early time and late time phases of our universe \cite{Bahamonde:2017ize}, to investigate whether the obtained solutions are stable or not. We will therefore construct the dynamical system of the present model in this subsection by defining the following dimensionless dynamical variables \cite{MW,Do:2021lyf,Do:2021pqk}
\begin{align}
X &= \frac{\dot\sigma}{\dot\alpha},\\
Y_1&=\frac{\dot\phi}{\dot\alpha},~Y_2 =\frac{\dot\psi}{\dot\alpha},\\
Z_1 &= \frac{p_A f_1}{\dot\alpha} e^{-\frac{1}{2}\left(a\phi +b\psi \right) -2\alpha-2\sigma},\\
Z_2& = \frac{q_A f_2}{\dot\alpha} e^{-\frac{1}{2}\left(c\phi +d\psi \right) -2\alpha-2\sigma},\\
W_1&= \frac{\sqrt{V_1}}{\dot\alpha},~W_2 =\frac{\sqrt{V_2}}{\dot\alpha},
\end{align}
with $V_1 \equiv V_{01}e^{\lambda_1 \phi}$ and $V_2 \equiv V_{02}e^{\lambda_2 \psi}$. As a result, the corresponding autonomous equations of the dynamical system can be defined from the field equations to be \cite{Do:2021lyf}
\begin{align}
\label{dynamical-1}
\frac{dX}{d\alpha} &= X \left[ 3\left(X^2-1 \right) +\frac{1}{2}\left(Y_1^2 +Y_2^2 \right)+\frac{1}{3}\left(Z_1^2+Z_2^2 \right) \right] +\frac{1}{3}\left(Z_1^2+Z_2^2 \right),\\
\frac{dY_1}{d\alpha} &= Y_1 \left[ 3\left(X^2-1 \right) +\frac{1}{2}\left(Y_1^2 +Y_2^2 \right)+\frac{1}{3}\left(Z_1^2+Z_2^2 \right) \right] +\frac{a}{2}Z_1^2 +\frac{c}{2}Z_2^2 -\lambda_1 W_1^2,\\
\frac{dY_2}{d\alpha} &= Y_2 \left[ 3\left(X^2-1 \right) +\frac{1}{2}\left(Y_1^2 +Y_2^2 \right)+\frac{1}{3}\left(Z_1^2+Z_2^2 \right) \right] +\frac{b}{2}Z_1^2 +\frac{d}{2}Z_2^2 -\lambda_2 W_2^2,\\
\frac{dZ_1}{d\alpha} &= Z_1 \left[ 3\left(X^2-1 \right) +\frac{1}{2}\left(Y_1^2+Y_2^2 \right)+\frac{1}{3} \left(Z_1^2+Z_2^2 \right) -2X-\frac{1}{2} \left(aY_1 +bY_2 \right)+1 \right],\\
\frac{dZ_2}{d\alpha} &= Z_2 \left[ 3\left(X^2-1 \right) +\frac{1}{2}\left(Y_1^2+Y_2^2 \right)+\frac{1}{3} \left(Z_1^2+Z_2^2 \right) -2X-\frac{1}{2} \left(cY_1 +dY_2 \right)+1 \right],\\
\frac{dW_1}{d\alpha} &=W_1 \left[ 3X^2 +\frac{1}{2} \left(Y_1^2+Y_2^2 \right) +\frac{1}{3} \left(Z_1^2+Z_2^2 \right) +\frac{\lambda_1}{2}Y_1 \right],\\
\frac{dW_2}{d\alpha} &=W_2 \left[ 3X^2 +\frac{1}{2} \left(Y_1^2+Y_2^2 \right) +\frac{1}{3} \left(Z_1^2+Z_2^2 \right) +\frac{\lambda_2}{2}Y_2 \right].
\end{align}
Here $\alpha \equiv \int \dot\alpha dt$ has been introduced as a dynamical time variable. It is apparent that the above dynamical system cannot be transformed from the field equations if the constraint equation derived from the Friedmann equation \eqref{field-equation-1},
\begin{equation} \label{constraint-dynamical}
\frac{V}{\dot\alpha^2} = W_1^2+W_2^2 =-3\left(X^2-1 \right)-\frac{1}{2}\left(Y_1^2+Y_2^2 \right) -\frac{1}{2} \left(Z_1^2+Z_2^2 \right),
\end{equation}
is ignored. Our next step will be finding out anisotropic fixed points ($X\neq 0$) to the dynamical system. To do this task, we will solve the following set of equations, 
\begin{equation}
\frac{dX}{d\alpha} =\frac{dY_1}{d\alpha}=\frac{dY_2}{d\alpha}=\frac{dZ_1}{d\alpha}=\frac{dZ_2}{d\alpha} =\frac{dW_1}{d\alpha}=\frac{dW_2}{d\alpha}=0.
\end{equation}
It is obvious from two equations, ${dW_1}/{d\alpha}={dW_2}/{d\alpha}=0$, that
\begin{align}\label{relation-1}
\lambda_1 Y_1 &=\lambda_2 Y_2,\\
\label{relation-2}
3X^2 +\frac{1}{2} \left(Y_1^2+Y_2^2 \right) +\frac{1}{3} Z^2+\frac{\lambda_1}{2}Y_1 &=0,
\end{align}
while two other equations, ${dZ_1}/{d\alpha}={dZ_2}/{d\alpha}=0$, give, with the help of Eq. \eqref{relation-2}, the following non-trivial relations,
\begin{align}\label{relation-3}
aY_1 +bY_2  =cY_1 +dY_2 &= \kappa_1 \lambda_1 Y_1,\\
\label{relation-4}
2X +\frac{1}{2}\lambda_1\left(\kappa_1+1  \right)Y_1+2 &=0,
\end{align}
 where $Z$ has been introduced as an additional variable, whose definition is given by  
 \begin{equation}
 Z^2= Z_1^2+Z_2^2.
 \end{equation}
It is noted that we have ignored the trivial solution, $Z_1=Z_2=W_1=W_2=0$, which is related to an isotropic fixed point with $X=0$. The remaining equations, i.e., ${dX}/{d\alpha}={dY_1}/{d\alpha}={dY_2}/{d\alpha}=0$, provide us with the corresponding ones,
 \begin{align}
 \label{relation-5}
 X \left(\frac{\lambda_1}{2}Y_1 +3 \right) -\frac{Z^2}{3} &=0,\\
 \label{relation-6}
 -\left(\frac{\lambda_1}{2}Y_1 +3 \right)Y_1 +\frac{a}{2} Z_1^2 +\frac{c}{2}Z_2^2 -\lambda_1 W_1^2 & =0,\\
  \label{relation-7}
 -\left(\frac{\lambda_2}{2}Y_2 +3 \right)Y_2 +\frac{b}{2} Z_1^2 +\frac{d}{2}Z_2^2 -\lambda_2 W_2^2 & =0.
 \end{align}
 Furthermore, multiplying both sides of Eq. \eqref{relation-6} with $\lambda_1^{-1}$ and both sides of Eq. \eqref{relation-7} with $\lambda_2^{-1}$ then combining the two obtained equations lead to
 \begin{equation}\label{relation-8}
-\left(\frac{\lambda_1}{2} Y_1 +3 \right) \left[ \left(\frac{1}{\lambda_1} +\frac{\lambda_1}{\lambda_2^2} \right)Y_1 +1 \right]+\left(\frac{\kappa_1}{2} +\frac{1}{6} \right)Z^2=0,
 \end{equation}
 with the help of other equations \eqref{constraint-dynamical}, \eqref{relation-1}, and \eqref{relation-2}.
It is apparent that we have addressed three equations, \eqref{relation-4}, \eqref{relation-5}, and \eqref{relation-8}, which are enough for solving three dynamical variables, $X$, $Y_1$, and $Z$. The other dynamical variable, $Y_2$, can be determined in terms of $Y_1$, according to the relation shown in Eq. \eqref{relation-1}.  As a result, a set of  non-trivial solutions of anisotropic fixed point can be defined from these equations to be
 \begin{align}
 X &= \frac{2 \left[ \lambda_1^2\lambda_2^2 \left(\kappa_1 +1 \right)-4\lambda_1^2-4 \lambda_2^2 \right]}{\lambda_1^2\lambda_2^2 \left(3\kappa_1^2 +4\kappa_1 +1 \right)+8\lambda_1^2+8\lambda_2^2 },\\
 Y_1&= \frac{-12 \lambda_1 \lambda_2^2 \left(\kappa_1+1 \right)}{\lambda_1^2\lambda_2^2 \left(3\kappa_1^2 +4\kappa_1 +1 \right)+8\lambda_1^2+8\lambda_2^2},\\
 Z^2 &= \frac{18 \left[ \lambda_1^2\lambda_2^2 \left(\kappa_1+1 \right)-4\lambda_1^2-4\lambda_2^2 \right] \left[ \lambda_1^2\lambda_2^2 \left(3\kappa_1^2+2\kappa_1 -1 \right) +8\lambda_1^2 +8\lambda_2^2 \right]}{\left[\lambda_1^2\lambda_2^2 \left(3\kappa_1^2 +4\kappa_1 +1 \right)+8\lambda_1^2+8\lambda_2^2\right]^2}.
 \end{align}
 One could ask if this anisotropic fixed point is equivalent to a set of anisotropic power-law solutions found in the previous section. If they are equivalent to each other, stability of the anisotropic fixed point will tell us that of anisotropic power-law solution. As a result, it is straightforward to verify that this anisotropic fixed point is absolutely equivalent to the derived anisotropic power-law solution. For instance, the equality $X =\eta/\zeta$ is obviously verified with $\zeta$ and $\eta$ defined in Eqs. \eqref{solution-zeta} and \eqref{solution-eta}, respectively.
 \subsection{Stability of anisotropic fixed point}
\begin{figure}[hbtp] 
\begin{center}
{\includegraphics[height=65mm]{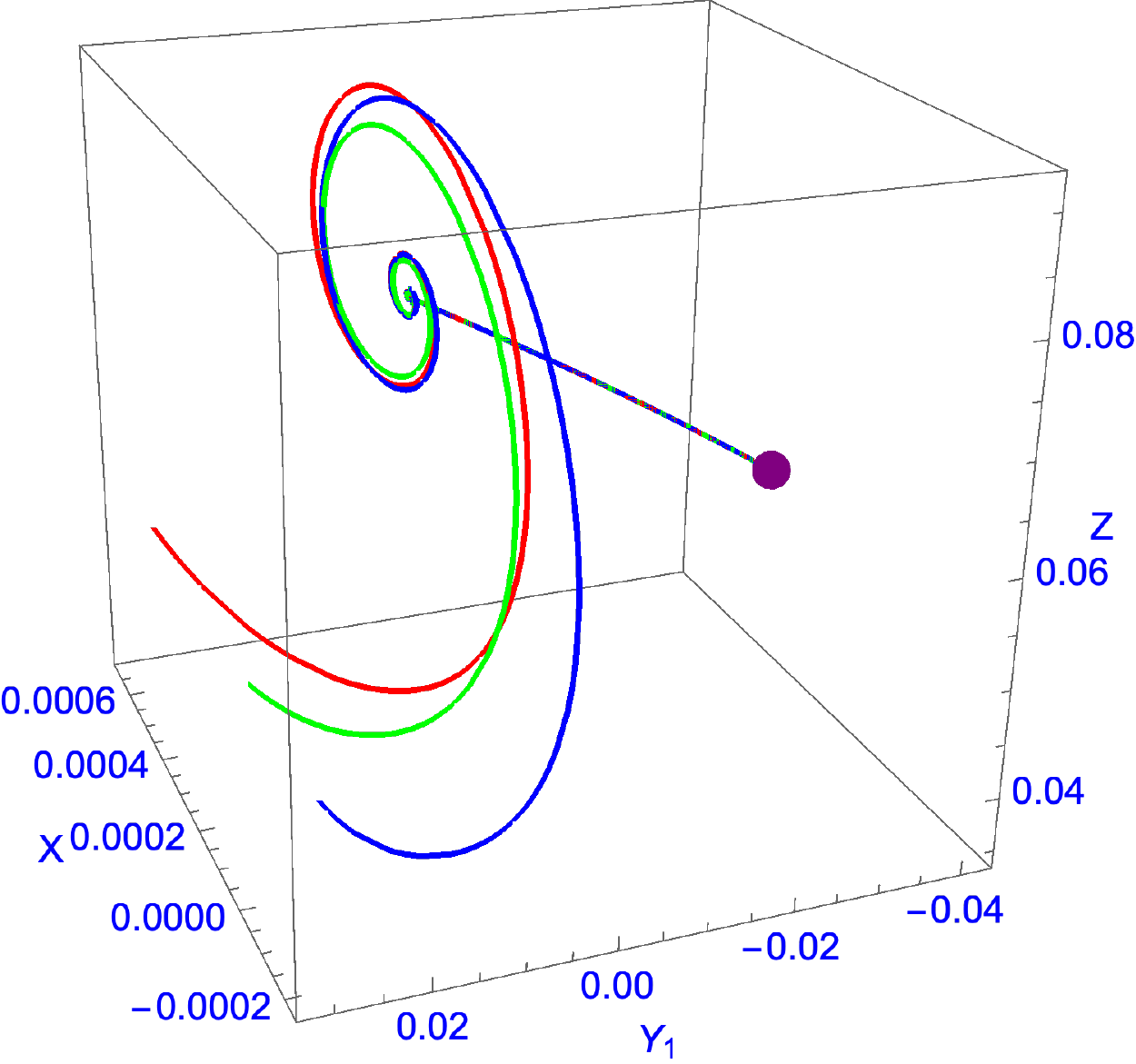}}\quad
{\includegraphics[height=65mm]{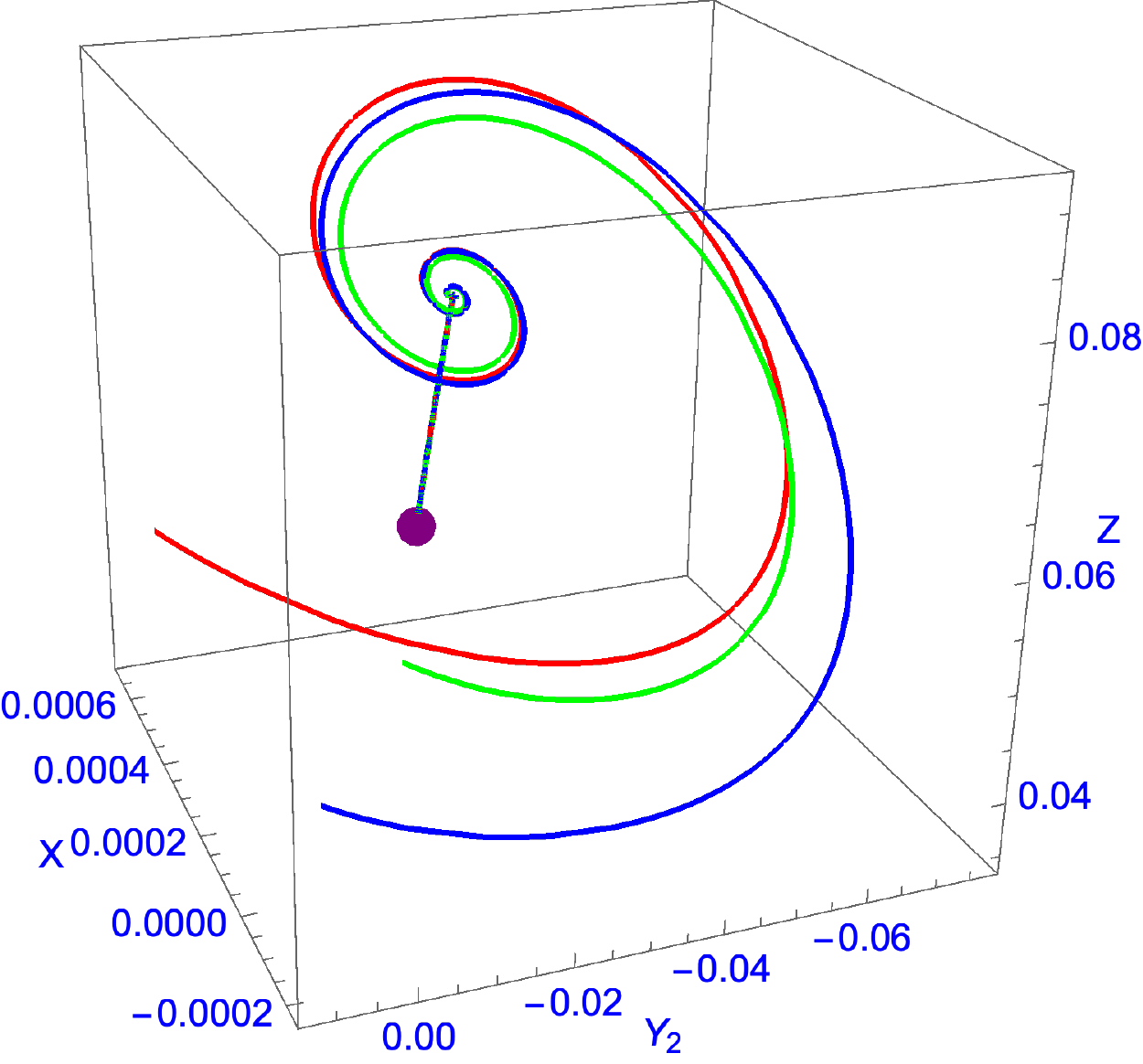}}\\
\caption{The anisotropic fixed point  (the purple points) as an attractor during the inflationary phase with parameters chosen as  $\lambda_1=0.1$, $\lambda_2=0.2$, $a=80$, $b=c=40$, and $d=120$.}
\label{fig1}
\end{center}
\end{figure}

\begin{figure}[hbtp] 
\begin{center}
{\includegraphics[height=65mm]{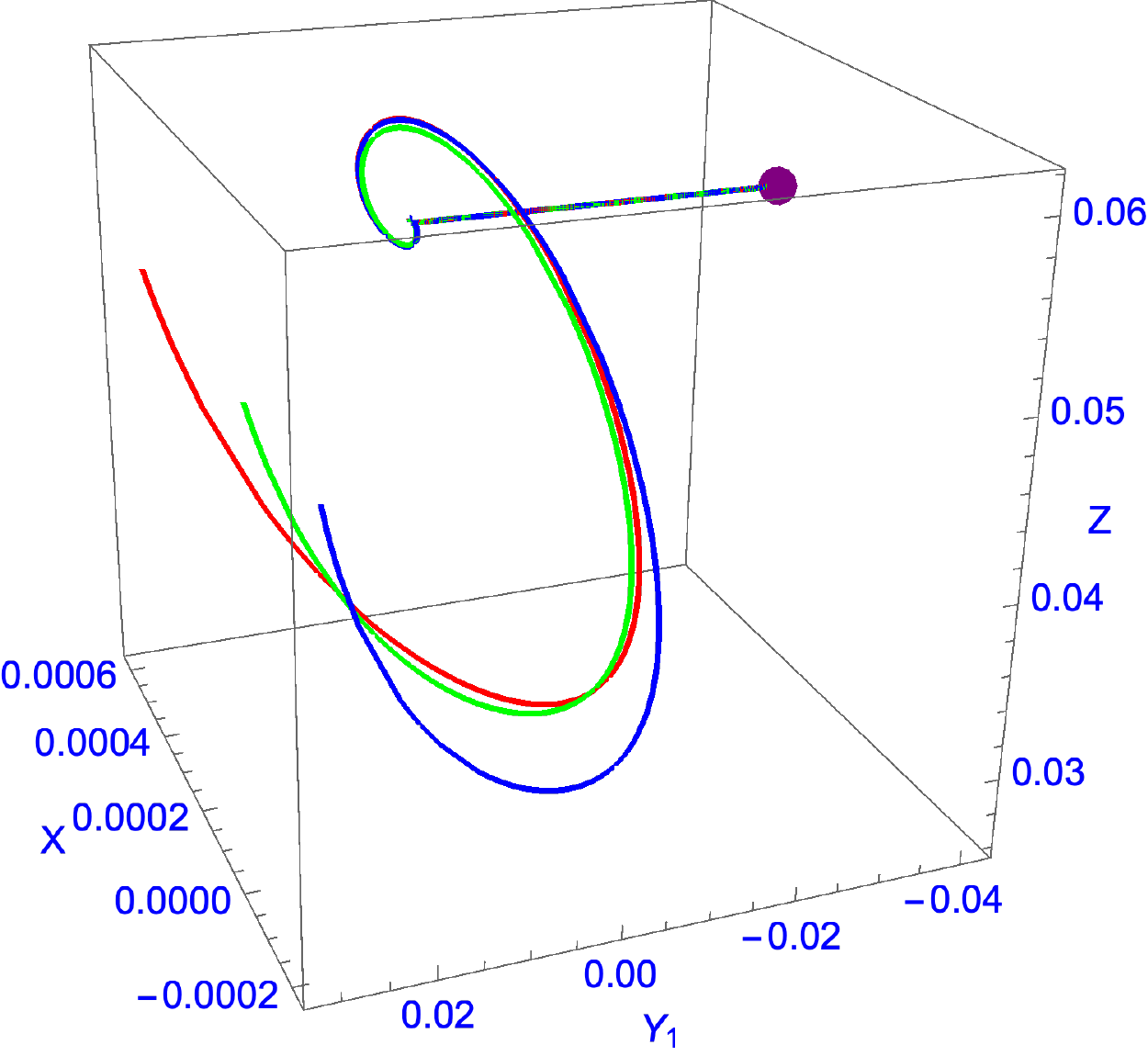}}\quad
{\includegraphics[height=65mm]{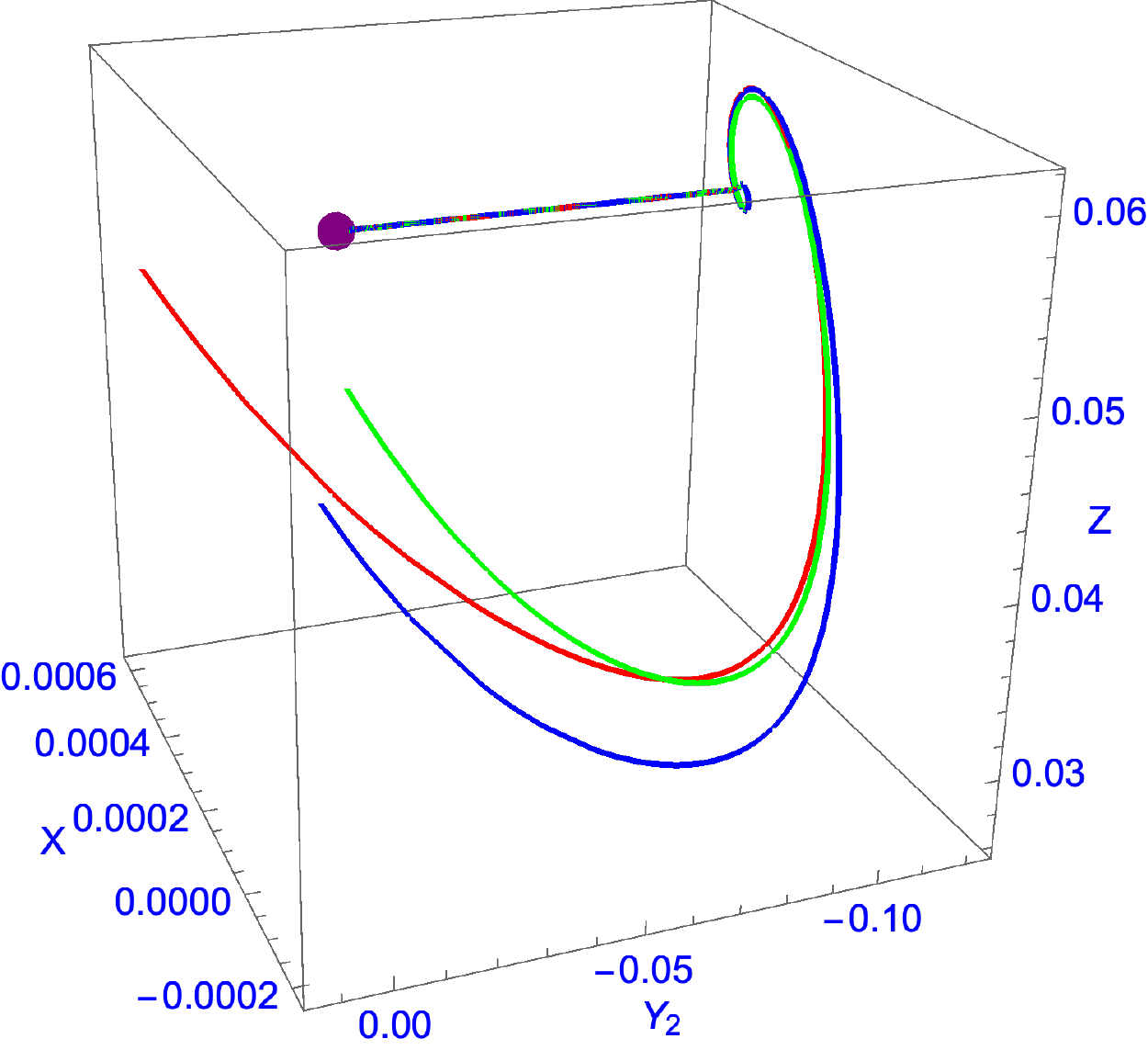}}\\
\caption{The anisotropic fixed point (the purple points) as an attractor during the inflationary phase with parameters chosen as  $\lambda_1=0.1$, $\lambda_2=0.2$, $a=c=80$, and $b=d=40$.}
\label{fig2}
\end{center}
\end{figure}

  During the inflationary phase with $\kappa_1 \gg 1$, $|\lambda_1| \ll 1$, and $|\lambda_2| \ll 1$, it is very useful to take the following approximations, 
 \begin{align}
 &X \simeq \frac{2}{3\kappa_1} \ll 1,\\
& Y_1 \simeq -\frac{4}{\lambda_1 \kappa_1} \ll 1,\\
& Y_2 \simeq -\frac{4}{\lambda_2 \kappa_1} \ll 1,\\
& Z^2 \simeq \frac{6}{\kappa_1} \ll 1,\\
& W_1^2+W_2^2 \simeq 3.
 \end{align}
 Perturbing the autonomous equations around the anisotropic fixed point will lead to the following perturbed equations,
 \begin{align}
\frac{d\delta X}{d\alpha} &\simeq -3\delta X,\\
\frac{d\delta Y_1}{d\alpha} &\simeq -3\delta Y_1  +aZ_1 \delta Z_1 +cZ_2 \delta Z_2 -2\lambda_1 W_1 \delta W_1,\\
\frac{d\delta Y_2}{d\alpha} &\simeq -3\delta Y_2  +bZ_1 \delta Z_1 +dZ_2 \delta Z_2 -2\lambda_2 W_2 \delta W_2,
\end{align}
\begin{align}
\frac{d\delta Z_1}{d\alpha} &\simeq -Z_1 \left[2\delta X +\frac{1}{2} \left(a\delta Y_1 + b\delta Y_2 \right) \right],\\
\frac{d\delta Z_2}{d\alpha} &\simeq -Z_2 \left[2\delta X +\frac{1}{2} \left(c\delta Y_1 + d\delta Y_2 \right) \right],\\
\frac{d\delta W_1}{d\alpha} &\simeq \frac{\lambda_1}{2}W_1 \delta Y_1,\\
\frac{d\delta W_2}{d\alpha} &\simeq \frac{\lambda_1}{2}W_2 \delta Y_2.
 \end{align}
 By considering exponential perturbations \cite{MW,Do:2021lyf},
 \begin{align} \label{expo-pertur}
& \delta X = A_1 e^{\tau \alpha},\\
& \delta Y_1 = A_2 e^{\tau\alpha}, ~\delta Y_2 =A_3 e^{\tau\alpha},\\
&\delta Z_1 =A_4 e^{\tau\alpha},~\delta Z_2 =A_5 e^{\tau\alpha},\\
& \delta W_1 =A_6e^{\tau\alpha},~\delta W_2 =A_7e^{\tau\alpha},
 \end{align}
 we can rewrite the above perturbed equations as a homogeneous matrix equation given by
 \begin{equation} \label{stability-equation}
{\cal M}\left( {\begin{array}{*{20}c}
   A_1  \\
   A_{2}  \\
   A_{3} \\
   A_{4}  \\
   A_{5}  \\
   A_{6}\\
   A_{7}\\
 \end{array} } \right) \equiv \left[ {\begin{array}{*{20}c}
   {-3-\tau} & {0} & {0 } & {0 } & {0} &{0} &{0} \\
   {0 } & {-3-\tau} & {0 } & {a Z_1} &{cZ_2}&{-2\lambda_1 W_1}&{0}  \\
     {0 } & {0} & {-3-\tau } & {bZ_1 } &{d  Z_2}&{0}&{-2\lambda_2 W_2}  \\
   {-2Z_1} & {-\frac{a}{2} Z_1 } & {-\frac{b}{2}Z_1 } & {-\tau } &{0}&{0}&{0} \\
   {-2Z_2} & {-\frac{c}{2} Z_2 } & {-\frac{d}{2} Z_2 } &{0}& {-\tau } &{0}&{0}\\
   {0}&{\frac{\lambda_1 }{2}W_1}&{0} &{0} &{0}&{-\tau}&{0}\\
{0}&{0}&{\frac{\lambda_2 }{2}W_2} &{0} &{0}&{0}&{-\tau}\\
 \end{array} } \right]\left( {\begin{array}{*{20}c}
    A_1  \\
   A_{2}  \\
   A_{3} \\
   A_{4}  \\
   A_{5}  \\
   A_{6}\\
   A_{7}\\
 \end{array} } \right) = 0.
\end{equation}
It is well known that non-trivial solutions of this equation exist only when 
\begin{equation}
\det {\cal M} =0,
\end{equation}
which can be defined to be an equation of $\tau$ such as
 \begin{equation} \label{poly}
 \tau^2 \left(\tau+3 \right) \left( a_4 \tau^4 +a_3\tau^3 + a_2 \tau^2+a_1 \tau +a_0 \right)=0,
 \end{equation}
 whose coefficients $a_i$ with $i$ runs from $0$ to $4$ are determined as
 \begin{align}
 a_4&=4>0,\\
 a_3&= 24>0,\\
 a_2&= 2 \left(a^2+b^2 \right)Z_1^2 +2 \left(c^2+d^2 \right)Z_2^2+4 \left(\lambda_1^2 W_1^2 +\lambda_2^2 W_2^2 \right)+36>0,\\
 a_1 &= 6 \left(a^2+b^2 \right)Z_1^2 +6 \left(c^2+d^2 \right)Z_2^2+12 \left(\lambda_1^2 W_1^2 +\lambda_2^2 W_2^2 \right)>0,\\
 a_0&= \left(bc-ad \right)^2 Z_1^2Z_2^2 +2\lambda_1^2 W_1^2 \left(b^2 Z_1^2 +d^2 Z_2^2 \right)+2\lambda_2^2 W_2^2 \left(a^2 Z_1^2+c^2Z_2^2 \right)+4\lambda_1^2 \lambda_2^2 W_1^2W_2^2 >0.
 \end{align}
 It should be noted that any positive roots $\tau>0$ will correspond to unstable modes since the corresponding exponential perturbations shown above will blow up when $\alpha$ becomes large. On the other hand, all non-positive roots $\tau \leq 0$ will correspond to stable modes since the corresponding exponential perturbations will tend to be suppressed when $\alpha$ becomes large.
 It turns out that Eq. \eqref{poly} will no longer admit any positive roots $\tau >0$ since all coefficients $a_i$ ($i=0-4$) are positive definite. This implies that the anisotropic fixed point is indeed stable against field perturbations. More interestingly, this result is supported by the attractor property of  the anisotropic fixed point confirmed by numerical calculations (see Figs. \ref{fig1} and \ref{fig2} for details). These results clearly demonstrate that the cosmic no-hair conjecture is still broken down  within the present model.  
\subsection{Effect of phantom field}
\begin{figure}[hbtp] 
\begin{center}
{\includegraphics[height=65mm]{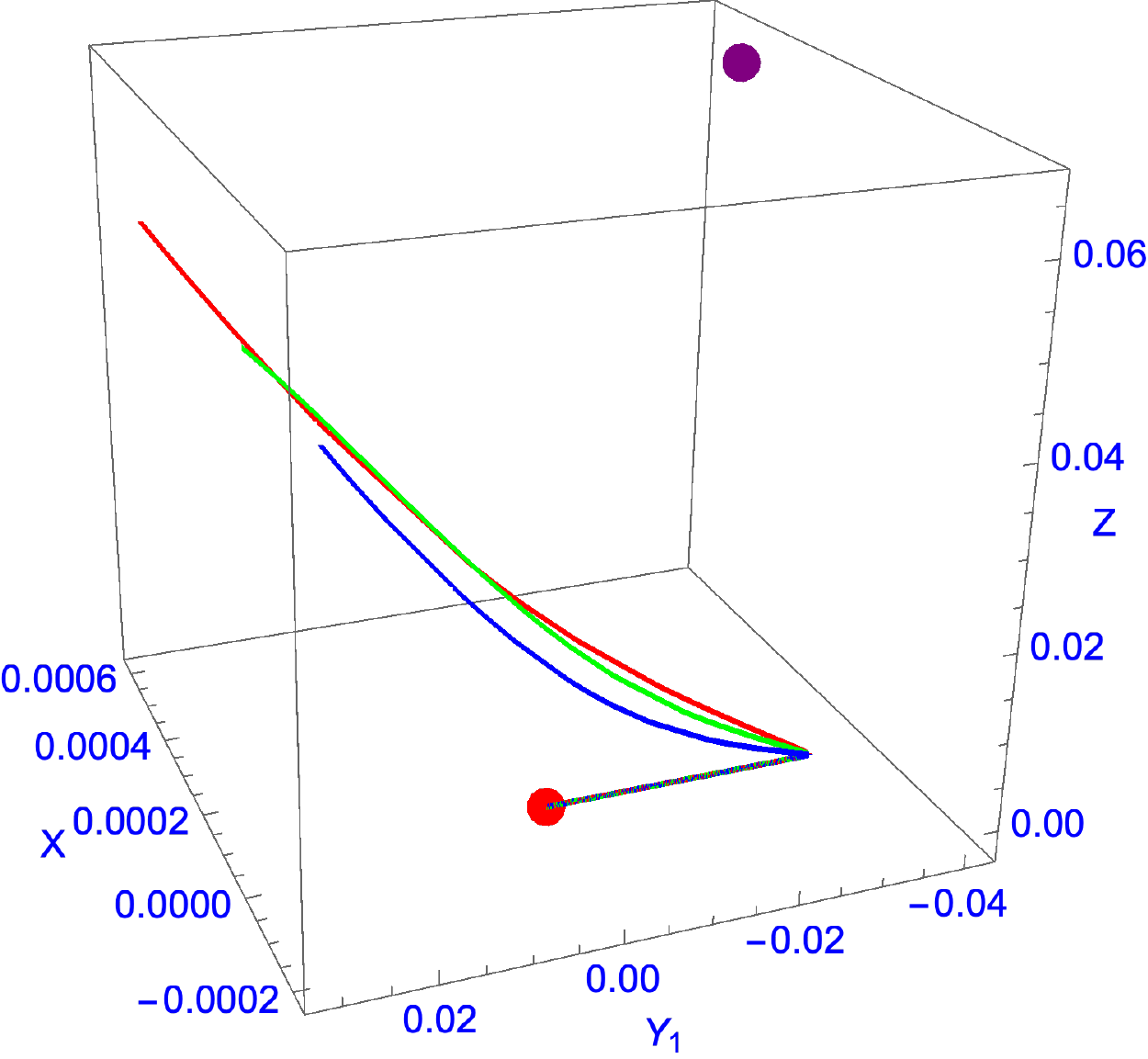}}\quad
{\includegraphics[height=65mm]{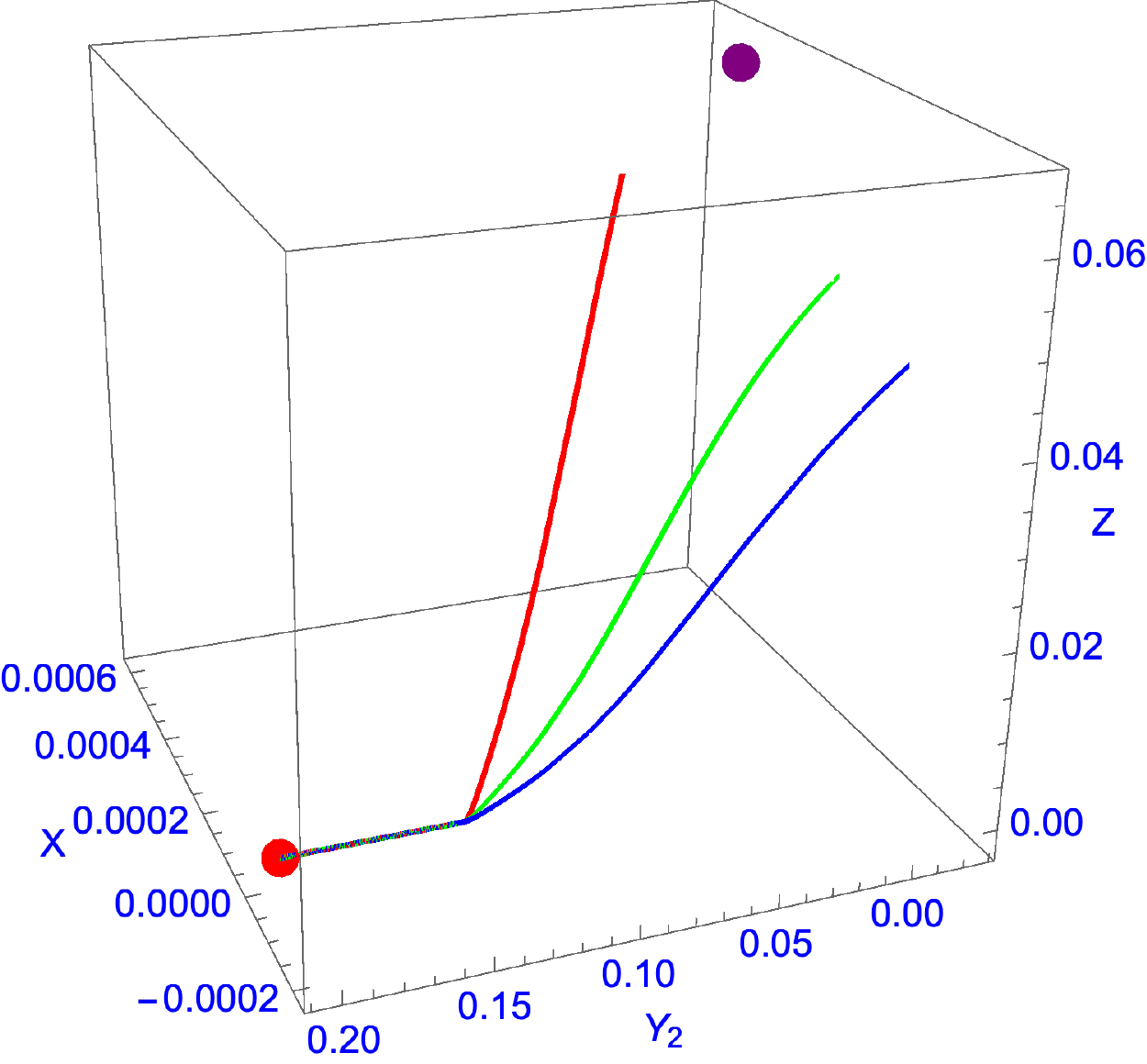}}\\
\caption{The isotropic fixed point (the red points) rather than the anisotropic fixed point (the purple points) is an attractor during the inflationary phase with parameters chosen as $\lambda_1=0.1$, $\lambda_2=0.2$, $a=80$, $b=c=40$, and $d=120$. }
\label{fig3}
\end{center}
\end{figure}

Motivated by our previous studies \cite{Do:2011zza,Do:2017rva,Do:2021lyf} we would like to see if a phantom field, whose kinetic term is negative definite \cite{Cai:2009zp,Cai:2008qb}, could save the cosmic no-hair conjecture by causing unstable mode(s) to its set of new anisotropic inflationary solutions. To do this, we will assume $\psi$ to be a phantom field, i.e.,  its kinetic term will be flipped from $- \partial_\mu \psi \partial^\mu \psi/2$ to $+\partial_\mu \psi \partial^\mu \psi/2$. Repeating all calculations, which have been done in the previous sections, for this phantom field, we are able to obtain a set of the corresponding solutions of $\zeta$ and $\eta$ as follows
\begin{align} \label{solution-zeta-phantom}
\zeta &=\frac{\lambda_1^2\lambda_2^2 \left(3\kappa_1^2 +4\kappa_1 +1 \right)-8\lambda_1^2+8\lambda_2^2  }{6\lambda_1^2 \lambda_2^2 \left(\kappa_1+1 \right) },\\
\eta & = \frac{ \lambda_1^2\lambda_2^2 \left(\kappa_1 +1 \right)+4\lambda_1^2-4 \lambda_2^2}{3\lambda_1^2 \lambda_2^2 \left(\kappa_1+1 \right) }.
\end{align}
Furthermore, the corresponding anisotropic fixed point will be solved to be
\begin{align}
X& = \frac{2 \left[ \lambda_1^2\lambda_2^2 \left(\kappa_1 +1 \right)+4\lambda_1^2-4 \lambda_2^2 \right]}{\lambda_1^2\lambda_2^2 \left(3\kappa_1^2 +4\kappa_1 +1 \right)-8\lambda_1^2+8\lambda_2^2 },\\
Y_1&= \frac{-12 \lambda_1 \lambda_2^2 \left(\kappa_1+1 \right)}{\lambda_1^2\lambda_2^2 \left(3\kappa_1^2 +4\kappa_1 +1 \right)-8\lambda_1^2+8\lambda_2^2},\\
 Z^2 &= \frac{18 \left[ \lambda_1^2\lambda_2^2 \left(\kappa_1+1 \right)+4\lambda_1^2-4\lambda_2^2 \right] \left[ \lambda_1^2\lambda_2^2 \left(3\kappa_1^2+2\kappa_1 -1 \right) -8\lambda_1^2 +8\lambda_2^2 \right]}{\left[\lambda_1^2\lambda_2^2 \left(3\kappa_1^2 +4\kappa_1 +1 \right)-8\lambda_1^2+8\lambda_2^2\right]^2}.
 \end{align}
It is clear that we still have the anisotropic power-law inflation even when $\psi$ is the phantom field if the constraints required in the previous sections are still valid in this case. Furthermore, if we turn off the second vector field $A^2_\mu$ and let $a=2\rho_1$ and $b=2\rho_2$, then a set of anisotropic inflationary solutions found in Ref. \cite{Do:2011zza} will be recovered, e.g.,
\begin{equation}
X = \frac{2\left[\lambda_1 \lambda_2 \left( \lambda_1 \lambda_2 +2 \lambda_1 \rho_2 +2\lambda_2 \rho_1 \right) +4\left(\lambda_1^2-\lambda_2^2 \right) \right] }{4\left( \lambda_1 \rho_2 +\lambda_2 \rho_1 \right) \left(2\lambda_1 \lambda_2 +3\lambda_1 \rho_2 +3\lambda_2 \rho_1 \right) +\lambda_1^2 \lambda_2^2 -8 \left(\lambda_1^2 -\lambda_2^2 \right)}.
\end{equation}
It should be noted that a set of anisotropic inflationary solutions found in Ref. \cite{Do:2011zza} has been shown to be unstable against power-law perturbations. To see if the set of anisotropic solutions obtained above is still unstable, we define its corresponding equation of $\tau$ to be
\begin{equation} \label{poly-phantom}
 \tau^2 \left(\tau+3 \right) \left( \bar a_4 \tau^4 +\bar a_3\tau^3 + \bar a_2 \tau^2+\bar a_1 \tau +\bar a_0 \right)=0,
 \end{equation}
 with
 \begin{align}
\bar a_4&=a_4>0,\\
 \bar a_0&=-a_0  <0.
 \end{align}
It turns out that the equation,
\begin{equation}
F(\tau)\equiv \bar a_4 \tau^4 +\bar a_3\tau^3 + \bar a_2 \tau^2+\bar a_1 \tau +\bar a_0 =0,
\end{equation}
will admit at least one positive root $\tau>0$ since $\bar a_0 \bar a_4 <0$. Indeed, due to the fact that $F(\tau=0)=\bar a_0 <0$ and $F(\tau \gg 1) \sim \bar a_4 \tau^4 >0$ and the curve $F(\tau)$ will therefore cross the positive horizontal $\tau$-axis at least one time at $\tau =\tau_\ast >0$. Mathematically, this crossing point $\tau =\tau_\ast$ is nothing but a positive root to the equation $F(\tau)=0$. The existence of this positive root will lead to a blowing up of perturbations when $\alpha$ become large and will therefore destroy the stability of the anisotropic fixed point.  Furthermore, numerical calculations also support this result by showing that the isotropic fixed point with $X=Z=0$ is indeed attractive in this case (see Fig. \ref{fig3} for details).
All these results clearly imply that the corresponding set of new anisotropic power-law inflationary solutions for the phantom field is indeed unstable. And the phantom field seems to favor the cosmic no-hair conjecture. Note that the inclusion of phantom would lead the corresponding quintom model to several interesting points, which single inflaton field models do not have, as summarized in the comprehensive review paper \cite{Cai:2009zp}. One of these points is that one may obtain a bouncing solution at early times of the universe  in a quintom scenario with a null energy condition violation. Therefore, this may provide a possible solution to the singularity problem of standard Big Bang cosmology \cite{Cai:2008qb}. For now, these points are beyond our current consideration associated with the stability of Bianchi type I inflationary model. However, we would revisit these points with the presence of vector fields in the future.  
\section{Conclusions} \label{final}
We have proposed a generalized model of two scalar and two vector fields, in which two scalar fields are allowed to simultaneously non-minimally couple to each vector field.  This is of course a non-trivial generalization of the previous model of two scalar and two vector fields \cite{Do:2021lyf}, in which one scalar field is only non-minimally coupled to one vector field. As a result, we have been able to derive the set of new exact anisotropic power-law inflationary solutions to the present model. It is more general than that obtained in the previous papers  \cite{MW,Do:2021lyf}. Stability analysis based on the dynamical system method has pointed out that this  set of new anisotropic power-law  inflationary solutions turns out to be stable and attractive as expected. This result emphasizes that the cosmic no-hair conjecture has one more counterexample coming from the generalized model of two scalar and two vector fields proposed in this paper. It supports an observation that  the non-minimal coupling terms between scalar and vector fields play the leading role in breaking down the validity of the cosmic no-hair conjecture. To support this conjecture, we have found that the phantom field seems to be a safeguard of the cosmic no-hair conjecture by making spatial anisotropies unstable and therefore diluted during the inflationary phase, similar to our previous papers  \cite{Do:2011zza,Do:2017rva,Do:2021lyf}.  Due to complicated and lengthy calculations, CMB imprints of multi-field extensions of the KSW model will be our future studies and will be published elsewhere. We hope that the present paper would be useful to studies on anisotropic inflation.
\begin{acknowledgments}
We would like to thank two anonymous referees for their useful comments and suggestions.  T.Q.D. was partially supported by the Vietnam National Foundation for Science and Technology Development (NAFOSTED) under grant number 103.01-2020.15. W.F.K. is supported in part by the Ministry of Science and Technology (MOST) of Taiwan under Contract No. MOST 110-2112-M-A49-007. 
\end{acknowledgments}

\end{document}